\documentclass[aps,prd,preprint,superscriptaddress]{revtex4}
\usepackage{graphicx}

\def\eg{{\it e.g.}}

\def\be{\begin{equation}}
\def\ee{\end{equation}}
\def\bea{\begin{eqnarray}}
\def\eea{\end{eqnarray}}
\def\bean{\begin{eqnarray*}}
\def\eean{\end{eqnarray*}}
\def\bary{\begin{array}}
\def\eary{\end{array}}

\def\bit{\begin{itemize}}
\def\eit{\end{itemize}}

\def\ol{\overline}

\def\cB{{\cal{B}}}

\def\vud{V_{ud}}
\def\vus{V_{us}}

\def\vtd{V_{td}}
\def\vts{V_{ts}}

\def\ub{{\bar u}}
\def\cb{{\bar c}}

\def\db{{\bar d}}
\def\sb{{\bar s}}
\def\bb{{\bar b}}

\begin{document}

\preprint{ANL-HEP-PR-01-119, EFI-01-56, hep-ph/0112285}
\preprint{December 2001}
\bigskip

\title{UPDATED ANALYSIS OF SOME TWO-BODY CHARMLESS $B$ DECAYS}

\author{Cheng-Wei Chiang}
\email[e-mail: ]{chengwei@hep.uchicago.edu}
\affiliation{HEP Division, Argonne National Laboratory
9700 S. Cass Avenue, Argonne, IL 60439}
\affiliation{Enrico Fermi Institute and Department of Physics,
University of Chicago, 5640 S. Ellis Avenue, Chicago, IL 60637}
\author{Jonathan L. Rosner}
\email[e-mail: ]{rosner@hep.uchicago.edu}
\affiliation{Enrico Fermi Institute and Department of Physics, 
University of Chicago, 5640 S. Ellis Avenue, Chicago, IL 60637}

\date{\today}

\begin{abstract}
  New data from the BaBar, Belle, and CLEO Collaborations on $B$ decays to
  two-body charmless final states are analyzed, with the following
  consequences: (1) The penguin amplitude which dominates the decay $B^+ \to
  \pi^+ K^{*0}$ has a magnitude similar to that dominating $B^+ \to \pi^+ K^0$.
  (2) The decay $B^+ \to \pi^+ \eta$, a good candidate for observing direct
  $CP$ violation, should be detectable at present levels of sensitivity. (3)
  The decays $B^+ \to \eta' K^+$ and $B^+ \to \eta K^{*+}$ are sufficiently
  similar in rate to the corresponding decays $B^0 \to \eta' K^0$ and $B^0 \to
  \eta K^{*0}$, respectively, that one cannot yet infer the need for ``tree''
  amplitudes $t'$ contributing to the $B^+$ but not the $B^0$ decays.
  Statistical requirements for observing this and other examples of
  tree-penguin interference are given. (4) Whereas the $B^+ \to \eta' K^+$ and
  $B^0 \to \eta' K^0$ rates cannot be accounted for by the penguin amplitude
  $p'$ alone but require an additional flavor-singlet penguin contribution
  $s'$, no such flavor-singlet penguin contribution is yet called for in the
  decays $B^+ \to \eta K^{*+}$ or $B^0 \to \eta K^{*0}$.  Predictions for the
  rates for $B^+ \to \eta' K^{*+}$ and $B^0 \to \eta' K^{*0}$ are given which
  would allow one to gauge the importance of these flavor-singlet penguin
  amplitudes.
\end{abstract}


\maketitle

\section{INTRODUCTION \label{sec:int}}

The decays of $B$ mesons are rich sources of information on fundamental aspects
of weak couplings as described by the Cabibbo-Kobayashi-Maskawa (CKM) matrix,
and on potential effects of physics beyond the Standard Model.  Especially
useful information can be obtained from $B$ decays to pairs of light charmless
mesons, both pseudoscalar ($P$) and vector ($V$).  A number of questions can
now be addressed more incisively in the light of recent data from the CLEO,
BaBar, and Belle detectors.  In the present paper we shall discuss several of
these, showing that progress is being made and setting goals of data samples
for more definitive answers.  We limit our discussion to a few topics.

(1) Recent measurements of the branching ratio for $B^+ \to \pi^+ K^{*0}$
indicate that the penguin amplitude dominating this decay has a magnitude not
too much smaller than that of the penguin amplitude dominating $B^+ \to \pi^+
K^0$.  We use this information, as well as new information on the decays $B \to
K (\rho,\omega,\phi)$, to discuss several open questions associated with penguin
contributions to $B \to PV$ decays. These include a conjectured relation
between two types of penguin amplitudes called $p'_P$ and $p'_V$ in Ref.\ 
\cite{VPUP} in which the spectator quark is incorporated into a pseudoscalar or
a vector meson, respectively.  Arguments first proposed by Lipkin \cite{HJLP}
suggest that such amplitudes would be equal and opposite.  The contribution of
electroweak penguin diagrams in suppressing the decays $B \to K (\omega,\phi)$
is also noted.

(2) In Refs.\ \cite{BSR} and \cite{Dighe:1995gq}, the decays $B^+ \to \pi^+
(\eta,\eta')$ were proposed as good candidates for detecting direct $CP$
violation.  Present data samples are approaching the sensitivity for observing
these modes, whose branching ratios are expected to be a few parts in $10^6$.
We update estimates for the branching ratio for these decays and indicate the
possible range of likely direct $CP$ asymmetries.

(3) It has been suggested by several sets of authors (see, e.g., Refs.\
\cite{VPUP,HHY,HSW,HY}) that the decays $B^+ \to \eta' K^+$ and $B^+ \to \eta
K^{*+}$ might be enhanced with respect to the corresponding decays $B^0 \to
\eta' K^0$ and $B^0 \to \eta K^{*0}$, respectively, as a consequence of
constructive interference between tree and penguin amplitudes.  We review this
suggestion in the light of the latest data and find that this conclusion is not
yet warranted.  We indicate the statistical precision that is likely to be
needed in order to establish tree-penguin interference in this and other
processes reliably.  For $B^{+,0}$ decays to charmless nonstrange final
states such interference involves the product $\cos \alpha \cos \delta$,
while for decays to charmless non-strange final states it involves
$\cos \gamma \cos \delta$, where $\alpha$ and $\gamma$ are weak phases of the
unitarity triangle, while $\delta$ is a relative strong phase between tree
and penguin amplitudes.

(4) Lipkin \cite{HJLP} has argued for the enhancement of the decays $B \to
\eta' K$ and $B \to \eta K^*$ as a result of constructive interference between
nonstrange and strange quark components of the $\eta'$ or $\eta$, and for the
suppression of the decays $B \to \eta K$ and $B \to \eta' K^*$ because of
correspondingly destructive interference.  However, an additional amplitude
associated with the flavor-singlet part of the $\eta$ and $\eta'$ is both
allowed \cite{Dighe:1995gq} and required for the proper description of the $B
\to \eta' K$ decay rates \cite{DGReta}.  The status of this amplitude, called
$s'$, is reviewed.  It is pointed out that it does not need to be as large as
the penguin amplitude $p'$ in order to explain the data if it interferes
constructively with $p'$.  At present, while no such singlet contribution is
needed to explain the data on $B \to \eta K^*$, the flavor-singlet component of
$\eta$ is small.  A much more incisive test would be available once the decays
$B \to \eta' K^*$ (both charged and neutral) are available, since the penguin
contributions of nonstrange and strange quarks in the $\eta'$ partially cancel
one another, while the flavor-singlet component of the $\eta'$ is dominant.
Predictions for these rates are given.

We discuss our notation in Section II.  Experimental data, their averages, and
the corresponding inputs to our determination of amplitudes are treated in
Section III.  We then discuss the above four questions in turn: penguin
contributions in $B \to PV$ decays (Section IV), direct $CP$ violation in $B \to
\pi^+ (\eta,\eta')$ (Section V), tree-penguin interference (Section VI), and
the role of the flavor-singlet amplitude (Section VII). We summarize in Section
VIII.  An Appendix contains details of decay constant calculations.

\section{NOTATION \label{sec:not}}

We use the following quark content and phase conventions:
\begin{itemize}
\item{ {\it Bottom mesons}: $B^0=d\bb$, ${\ol B}^0=b\db$, $B^+=u\bb$,
    $B^-=-b\ub$, $B_s=s\bb$, ${\ol B}_s=b\sb$;}
\item{ {\it Charmed mesons}: $D^0=-c\ub$, ${\ol D}^0=u\cb$, $D^+=c\db$,
    $D^-=d\cb$, $D_s^+=c\sb$, $D_s^-=s\cb$;}
\item{ {\it Pseudoscalar mesons}: $\pi^+=u\db$, $\pi^0=(d\db-u\ub)/\sqrt{2}$,
    $\pi^-=-d\ub$, $K^+=u\sb$, $K^0=d\sb$, ${\ol K}^0=s\db$, $K^-=-s\ub$,
    $\eta=(s\sb-u\ub-d\db)/\sqrt{3}$,
    $\eta^{\prime}=(u\ub+d\db+2s\sb)/\sqrt{6}$;}
\item{ {\it Vector mesons}: $\rho^+=u\db$, $\rho^0=(d\db-u\ub)/\sqrt{2}$,
    $\rho^-=-d\ub$, $\omega=(u\ub+d\db)/\sqrt{2}$, $K^{*+}=u\sb$,
    $K^{*0}=d\sb$, ${\ol K}^{*0}=s\db$, $K^{*-}=-s\ub$, $\phi=s\sb$.}
\end{itemize}

In the present approximation there are seven types of independent amplitudes: a
``tree'' contribution $t$; a ``color-suppressed'' contribution $c$; a
``penguin'' contribution $p$; a ``singlet penguin'' contribution $s$, in which
a color-singlet $q \bar q$ pair produced by two or more gluons or by a $Z$ or
$\gamma$ forms an SU(3) singlet state; an ``exchange'' contribution $e$, an
``annihilation'' contribution $a$, and a ``penguin annihilation'' contribution
$pa$.  These amplitudes contain both the leading-order and electroweak penguin
contributions:
\be\bary{lll}
\label{eqn:dict}
t \equiv T + P_{EW}^C ~, &\quad& c \equiv C + P_{EW} ~, \\
p \equiv P - \frac{1}{3} P_{EW}^C ~, &\quad&
s \equiv S - \frac{1}{3} P_{EW} ~, \\
a \equiv A ~, &\quad& e + pa \equiv E + PA ~,
\eary\ee
where the capital letters denote the leading-order contributions
(\cite{Dighe:1995gq,Gronau:1994rj,Gronau:1995hn}) while $P_{EW}$ and $P_{EW}^c$
are respectively color-favored and color-suppressed electroweak penguin
amplitudes \cite{Gronau:1995hn}.  We shall neglect smaller terms
\cite{EWVP,GR2001} $P^E_{EW}$ and $P^A_{EW}$ [$(\gamma,Z)$-exchange and
$(\gamma,Z)$-direct-channel electroweak penguin amplitudes].  We shall denote
$\Delta S = 0$ transitions by unprimed quantities and $|\Delta S| = 1$
transitions by primed quantities.  For $PV$ decay modes, the subscript $P$ or
$V$ denotes the final-state meson (pseudoscalar or vector) incorporating the
spectator quark.  Although one $B \to VV$ decay ($B^0 \to \phi K^{*0}$) has
been seen, we shall not discuss such processes further here.

For the ${\ol b} \to {\ol d}$ and ${\ol b} \to {\ol u} u {\ol d}$ transitions,
an educated guess of the hierarchies among the amplitudes \cite{Gronau:1995hn}
is given in Table \ref{tab:hier}.  One notices that for $|\Delta S| = 1$
transitions, $c'$ contains an electroweak penguin amplitude at the next order.
Therefore, we put $c'$ together with $t'$ at the same order.  Similarly, since
part of the singlet amplitude is the electroweak penguin, $s'$ is at least of
order $P'_{EW}$.

%
% This is Table I
\begin{table}[t]
\caption{Hierarchies among magnitudes of flavor-$SU(3)$ amplitudes in powers of
 a parameter $\lambda \equiv |\vus| \simeq 0.22$.
\label{tab:hier}}
\begin{center}
\begin{tabular}{cccccc}
\hline\hline
  & $O(1)$ & $O(\lambda)$ & $O(\lambda^2)$ 
       & $O(\lambda^3)$ & $O(\lambda^4)$ \\
\hline
$\Delta S=0$
  & $T$ & $C,P$ & $E,A,P_{EW}$ 
       & $PA,P_{EW}^C$ & ~$PA_{EW}$~ \\
  & $t$ & $c,p$ & $e,a,s$ & $pa$ & \\
\hline
$|\Delta S|=1$
  & $P'$ & ~$T',P'_{EW}$~ & $C',PA',P_{EW}^{\prime C}$ 
       & ~$E',A',PA'_{EW}$~ & \\
  & $p'$ & $t',c',s'$ & $pa'$ & $e',a'$ & \\
\hline\hline

\end{tabular}
\end{center}
\vspace{0.5cm}
\end{table}

\section{AMPLITUDE DECOMPOSITIONS AND EXPERIMENTAL RATES \label{sec:amp}}

We list theoretical predictions and averaged experimental data for interesting
charmless $B$ decays involving $\Delta S = 0$ transitions in Table
\ref{tab:dS0} and those involving $|\Delta S| = 1$ transitions in Table
\ref{tab:dS1}.  Amplitudes of order $\lambda^2$ and smaller in Table
\ref{tab:hier} are omitted unless dominant.  Detailed experimental values are
listed in Tables \ref{tab:dS0data} and \ref{tab:dS1data}.  We will assume
\cite{VPUP} $p_V = - p_P$ and $p'_V = - p'_P$.  The averaged rates are obtained
by combining the data recently reported from CLEO, BaBar, and Belle groups
\cite{Jessop:1999cv,Cronin-Hennessy:kg,Richichi:1999kj,Jessop:2000bv,%
Briere:2001ue,Asner:2001eh,Gao:2001ce,Aubert:2000vr,Aubert:2001zd,%
Aubert:2001hs,Aubert:2001ye,Aubert:2001zf,Aubert:2001ap,Dallapiccola,%
Abe:2001nq,Bozek:2001xd,Abe:2001pf,Tajima:2001qp,belle0115,belle0137}.  In this
section we shall comment on some of the methods used to determine the invariant
amplitudes, deferring discussions of others to subsequent sections.

%
% This is Table II
\begin{table}[t]
\caption{Summary of predicted contributions to selected $\Delta S = 0$
decays of $B$ mesons.  Branching ratios ($\cB$) are quoted in units of
$10^{-6}$.  Numbers in italics are assumed inputs.  Experimental values
are averaged over results in Refs.\
\cite{Jessop:1999cv,Cronin-Hennessy:kg,Richichi:1999kj,Jessop:2000bv,%
Briere:2001ue,Asner:2001eh,Gao:2001ce,Aubert:2000vr,Aubert:2001zd,%
Aubert:2001hs,Aubert:2001ye,Aubert:2001zf,Aubert:2001ap,Dallapiccola,%
Abe:2001nq,Bozek:2001xd,Abe:2001pf,Tajima:2001qp,belle0115,belle0137}.
\label{tab:dS0}}
\begin{center}
\begin{tabular}{llcccccc}
\hline\hline
 & Mode & Amplitudes & $|t(+c)|^2$ & $|p|^2$ 
 & $|s|^2$ (a) & $|s|^2$ (b) & Expt. \\ 
\hline
$B^+ \to$
    & $\pi^+\pi^0$ & $-\frac{1}{\sqrt{2}}(t+c)$
        & 4.7 & 0 & 0 & 0 & $5.7 \pm 1.5$ \\
    & $K^+\ol{K}^0$ & $p$
        & 0 & 0.55 & 0 & 0 & $<2.4$ \\
    & $\pi^+\eta$ & $-\frac{1}{\sqrt{3}}(t+c+2p+s)$ 
        & 3.1 & 0.73 & 0.04 & 0.18 & $ < 5.7$ \\
    & $\pi^+\eta'$ & $\frac{1}{\sqrt{6}}(t+c+2p+4s)$ 
        & 1.6 & 0.37 & 0.35 & 1.4 & $<7$ \\
    & $\pi^+\rho^0$ 
        & $-\frac{1}{\sqrt{2}}(t_V+c_P+p_V-p_P)$
        & 7.9 & 0.78 & 0 & 0 & $12.8 \pm 3.6$ \\
%    & $\pi^0\rho^+$ & $-\frac{1}{\sqrt{2}}(t_P+c_V+p_P-p_V)$
%        & - & - & - & - & - \\
    & $\pi^+\omega$ 
        & $\frac{1}{\sqrt{2}}(t_V+c_P+p_P+p_V+2s_P)$
        & \textit{7.9} (c) & $\simeq 0$ & $\sim 0.01$ (d) 
        & - & $7.9 \pm 1.8$ \\
    & $\pi^+ \phi$
        & $s_P$ & 0 & 0 & 0.02 & - & $<1.4$ \\
\hline
$B^0 \to$
    & $\pi^+\pi^-$ & $-(t+p)$ 
        & {\it 7.3} &  0.51 & 0 & 0 & $4.4 \pm 0.9$ \\
    & $\pi^0\pi^0$ & $-\frac{1}{\sqrt{2}}(c-p)$ 
        & 0.04 & 0.26 & 0 & 0 & $<5.7$ \\
    & $K^+K^-$ & $-(e+pa)$ 
        & 0 & 0 & 0 & 0 & $<1.9$ \\
    & $\pi^{\pm}\rho^{\mp}$ 
        & $-(t_{(V,P)}+p_{(V,P)})$ 
        & 14.7 (e) & 0.36 (f) & 0 & 0 & $25.8 \pm 4.5$ (g) \\
    & $\pi^0\omega$ 
        & $\frac{1}{2}(c_P-c_V+p_P+p_V+2s_P)$
        & - & $\simeq 0$ & $ <   0.01$ (d) & - & $<3$ \\
\hline\hline
\end{tabular}
\end{center}
\leftline{(a) Assuming constructive interference between $s'$ and $p'$
in $B \to \eta' K$ (Table III).}
\leftline{(b) Assuming no interference between $s'$ and $p'$
in $B \to \eta' K$ (Table III).}
\leftline{(c) Neglecting other contributions to decay rate.}
\leftline{(d) $(c_P + 2 s_P)/\sqrt{2}$ contributes a term $\frac{1}{3}
  P_{EWP}/\sqrt{2}$ to amplitude.}
\leftline{(e) $|t_V|^2 = 14.7 \pm 3.3$ contribution to $\cB(B^0 \to \pi^+
\rho^-)$ estimated from $B^+ \to \pi^+ \omega$, neglecting}
\leftline{\qquad $c_P$ and $s_P$, leaving $|t_P|^2 = 11.1 \pm 5.6$ contributing
  to $\cB(B^0 \to \pi^- \rho^+)$.}
\leftline{(f) $|p_P|^2$ contribution to $\cB(B^0 \to \pi^- \rho^+)$ and
$|p_V|^2$ contribution to $\cB(B^0 \to \pi^+ \rho^-)$.}
\leftline{(g) Combined branching ratio for $\pi^+ \rho^-$ and $\pi^- \rho^+$.}
\end{table}
%

%
% This is Table III.
\begin{table}[t]
\caption{Same as Table II for $|\Delta S| = 1$ decays of $B$ mesons.
\label{tab:dS1}}
\begin{center}
\begin{tabular}{llccccccc}
\hline\hline
 & Mode & Amplitudes & $|t'|^2$ & $|p'|^2$ 
 & $|s'|^2$ (a) & $|s'|^2$ (b) & Expt. \\ 
\hline
$B^+ \to$
    & $\pi^+K^0$ & $p'$ 
        & 0 & {\it 17.2} & 0 & 0 & $17.2\pm2.6$ \\
    & $\pi^0K^+$ & $-\frac{1}{\sqrt{2}}(p'+t'+c')$ 
        & $0.30$ & 8.6 & 0 & 0 & $12.0\pm1.6$ \\
    & $\eta K^+$ & $-\frac{1}{\sqrt{3}}(t'+c'+s')$ 
        & $0.20$ & 0 & 1.4 & 5.6 & $<6.9$ \\
    & $\eta'K^+$ & $\frac{1}{\sqrt{6}}(3p'+t'+c'+4s')$ 
        & $0.10$ & 25.9 & 10.9 & 44.4 & $75\pm7$ \\
    & $\pi^+K^{*0}$ & $p'_P$
        & 0 & {\it 12.2} & 0 & 0 & $12.2\pm2.4$ \\
    & $\eta K^{*+}$ 
        & $-\frac{1}{\sqrt{3}}(p'_P-p'_V+t'_P+c'_V+s'_V)$
        & 0.22 & 16.2 & - & - & $24.5\pm7.1$ \\
    & $\eta'K^{*+}$ 
        & $\frac{1}{\sqrt{6}}(p'_P+2p'_V+t'_P+c'_V+4s'_V)$
        & 0.11 & 2.0 & - & - & $<35$ \\
    & $K^+ \omega$ 
        & $\frac{1}{\sqrt{2}}(p'_V+t'_V+c'_P+2s'_P)$ 
        & 0.60 & 6.1 & 0.24 (c) & - & $<4$ \\
    & $K^+ \phi$ & $p'_P+s'_P$ 
        & 0 & 12.2 & 0.48 & - & $7.7\pm1.2$ \\
\hline
$B^0 \to$
    & $\pi^-K^+$ & $-(p'+t')$ 
        & $0.56$ & 16.1 & 0 & 0 & $17.3\pm1.5$ \\
    & $\pi^0K^0$ & $\frac{1}{\sqrt{2}}(p'-c')$ 
        & 0 & 8.1 & 0 & 0 & $10.4\pm2.6$ \\
    & $\eta K^0$ & $-\frac{1}{\sqrt{3}}(c'+s')$ 
        & 0 & 0 & 1.3 & 5.2 & $<9.3$ \\
    & $\eta'K^0$ & $\frac{1}{\sqrt{6}}(3p'+c'+4s')$ 
        & 0 & 24.2 & 10.2 & 41.6 & $56\pm9$ \\
    & $\pi^-K^{*+}$ & $-(p'_P+t'_P)$ 
        & 0.62 & 11.4 & 0 & 0 & $23.8\pm6.1$ \\
    & $\eta K^{*0}$ & $-\frac{1}{\sqrt{3}}(p'_P-p'_V+c'_V+s'_V)$
        & 0 & 15.2 & - & - & $18.0\pm3.2$ \\
    & $\eta'K^{*0}$ & $\frac{1}{\sqrt{6}}(p'_P+2p'_V+c'_V+4s'_V)$ 
        & 0 & 1.9 & - & - & $<24$ \\
    & $K^+ \rho^-$ & $-(p'_V+t'_V)$
        & 1.13 & 11.4 & 0 & 0 & $15.9\pm4.4$ \\
    & $K^0 \omega$ & $\frac{1}{\sqrt{2}}(p'_V+c'_P+2s'_P)$
        & 0 & 5.7 & 0.23 (c) & - & $<13$ \\
    & $K^0 \phi$ & $p'_P+s'_P$ 
        & 0 & 11.4 & 0.45 & - & $7.5\pm1.8$ \\
\hline\hline
\end{tabular}
\end{center}
\leftline{(a): Maximal interference between $p'$ and $s'$ amplitudes assumed:
  constructive for $\eta K$ and $\eta' K$;}
\leftline{\qquad destructive for $K \phi$.}
\leftline{(b): No interference between $p'$ and $s'$ amplitudes assumed.}
\leftline{(c): $(c'_P+2s'_P)/\sqrt{2}$ contributes a term
$\frac{1}{3}P'_{EWP}/\sqrt{2} \simeq - 0.20 p'_V/\sqrt{2}$ to amplitude.}
\end{table}
%

% This is Table IV.
\begin{table}[t]
\caption{Experimental branching ratios of selected $\Delta S = 0$
decays of $B$ mesons.  Branching ratios are quoted in units of
$10^{-6}$.  Numbers in parentheses are upper bounds at 90 \% c.l.
References are given in square brackets.
\label{tab:dS0data}}
\begin{center}
\begin{tabular}{lllll}
\hline\hline
 & Mode & CLEO & BaBar & Belle \\ 
\hline
$B^+ \to$
    & $\pi^+\pi^0$ 
        & $5.6^{+2.6}_{-2.3}\pm1.7 \; (<12.7)$ \cite{Cronin-Hennessy:kg}
        & $5.1^{+2.0}_{-1.8}\pm0.8 \; (<9.6)$ \cite{Aubert:2001hs}
        & $7.8^{+3.8+0.8}_{-3.2-1.2} \; (<13.4)$ \cite{Abe:2001nq} \\
    & $K^+\ol{K}^0$ 
        & $<5.1$ \cite{Cronin-Hennessy:kg}
        & $-1.3^{+1.4}_{-1.0}\pm0.7 \; (<2.4)$ \cite{Aubert:2001hs}
        & $<5.0$ \cite{Abe:2001nq} \\
    & $\pi^+\eta$ 
        & $1.2^{+2.8}_{-1.2} \; (<5.7)$ \cite{Richichi:1999kj}
        & - & - \\
    & $\pi^+\eta'$ 
        & $1.0^{+5.8}_{-1.0} \; (<12)$ \cite{Richichi:1999kj}
        & $5.4^{+3.5}_{-2.6}\pm0.8 \; (<12)$ \cite{Aubert:2001zf}
        & $<7$ \cite{Abe:2001pf} \\
    & $\pi^+\rho^0$ 
        & $10.4^{+3.3}_{-3.4}\pm2.1$ \cite{Jessop:2000bv}
        & $24\pm8\pm3$ \cite{Aubert:2000vr}
        & $<14.5$ \cite{belle0115} \\
    & $\pi^+\omega$ 
        & $11.3^{+3.3}_{-2.9}\pm1.4$ \cite{Jessop:2000bv}
        & $6.6^{+2.1}_{-1.8}\pm0.7$ \cite{Aubert:2001zf}
        & $<9.4$ \cite{Bozek:2001xd} \\
    & $\pi^+ \phi$ 
        & - 
        & $0.21^{+0.49}_{-0.21}\pm0.05 \; (<1.4)$ \cite{Aubert:2001zd}
        & - \\
\hline
$B^0 \to$
    & $\pi^+\pi^-$ 
        & $4.3^{+1.6}_{-1.4}\pm0.5$ \cite{Cronin-Hennessy:kg}
        & $4.1\pm1.0\pm0.7$ \cite{Aubert:2001hs}
        & $5.6^{+2.3+0.4}_{-2.0-0.5}$ \cite{Abe:2001nq} \\
    & $\pi^0\pi^0$ 
        & $2.2^{+1.7+0.7}_{-1.3-0.7} \; (<5.7)$ \cite{Asner:2001eh} 
        & - & - \\
    & $K^+K^-$ 
        & $<1.9$ \cite{Cronin-Hennessy:kg}
        & $0.85^{+0.81}_{-0.66}\pm0.37 \; (<2.5)$ \cite{Aubert:2001hs}
        & $<2.7$ \cite{Abe:2001nq} \\
    & $\pi^{\pm}\rho^{\mp}$ 
        & $27.6^{+8.4}_{-7.4}\pm4.2$ \cite{Jessop:2000bv}
        & $28.9 \pm 5.4 \pm 4.3$ \cite{Dallapiccola}
        & $20.2^{+8.3}_{-6.6}\pm3.3 \; (<35.7)$ \cite{Bozek:2001xd} \\
    & $\pi^0\omega$ 
        & $0.8^{+1.9+1.0}_{-0.8-0.8} \; (<5.5)$ \cite{Jessop:2000bv} 
        & $-0.3\pm1.1\pm0.3 \; (<3)$ \cite{Aubert:2001zf}
        & - \\
\hline\hline
\end{tabular}
\end{center}
\vspace{0.5cm}
\end{table}
%

% This is Table V.
\begin{table}[t]
\caption{Same as Table IV for $|\Delta S| = 1$ decays of $B$ mesons.
\label{tab:dS1data}}
\begin{center}
\begin{tabular}{lllll}
\hline\hline
 & Mode & CLEO & BaBar & Belle \\ 
\hline
$B^+ \to$
    & $\pi^+K^0$ 
        & $18.2^{+4.6}_{-4.0}\pm1.6$ \cite{Cronin-Hennessy:kg}
        & $18.2^{+3.3}_{-3.0}\pm2.0$ \cite{Aubert:2001hs}
        & $13.7^{+5.7+1.9}_{-4.8-1.8}$ \cite{Abe:2001nq} \\
    & $\pi^0K^+$ 
        & $11.6^{+3.0+1.4}_{-2.7-1.3}$ \cite{Cronin-Hennessy:kg}
        & $10.8^{+2.1}_{-1.9}\pm1.0$ \cite{Aubert:2001hs}
        & $16.3^{+3.5+1.6}_{-3.3-1.8}$ \cite{Abe:2001nq} \\
    & $\eta K^+$ 
        & $2.2^{+2.8}_{-2.2} \; (<6.9)$ \cite{Richichi:1999kj}
        & - & - \\
    & $\eta'K^+$ 
        & $80^{+10}_{-9}\pm7$ \cite{Richichi:1999kj}
        & $70\pm8\pm5$ \cite{Aubert:2001zf}
        & $79^{+12}_{-11}\pm9$ \cite{Abe:2001pf} \\
    & $\pi^+K^{*0}$ 
        & $7.6^{+3.5}_{-3.0}\pm1.6 \; (<16)$ \cite{Jessop:2000bv}
        & $15.5 \pm 3.4 \pm 1.8$ \cite{Aubert:2001ap} 
        & $16.7^{+3.7+2.1+3.0}_{-3.4-2.1-5.9}$ \cite{belle0115} \\
    & $\eta K^{*+}$ 
        & $26.4^{+9.6}_{-8.2}\pm3.3$ \cite{Richichi:1999kj}
        & $22.1^{+11.1}_{-9.2}\pm3.3 \; (<33.9)$ \cite{Aubert:2001ye}
        & $<49.9$ \cite{belle0137} \\
    & $\eta'K^{*+}$ 
        & $11.1^{+12.7}_{-8.0} \; (<35)$ \cite{Richichi:1999kj} 
        & - & - \\
    & $K^+ \omega$ 
        & $3.2^{+2.4}_{-1.9}\pm0.8 \; (<7.9)$  \cite{Jessop:2000bv}
        & $1.4^{+1.3}_{-1.0}\pm0.3 \; (<4)$ \cite{Aubert:2001zf}
        & $<10.5$ \cite{Bozek:2001xd} \\
    & $K^+ \phi$ 
        & $5.5^{+2.1}_{-1.8}\pm0.6$ \cite{Briere:2001ue}
        & $7.7^{+1.6}_{-1.4}\pm0.8$ \cite{Aubert:2001zd}
        & $11.2^{+2.2}_{-2.0}\pm1.4$ \cite{Tajima:2001qp} \\
\hline
$B^0 \to$
    & $\pi^-K^+$ 
        & $17.2^{+2.5}_{-2.4}\pm1.2$ \cite{Cronin-Hennessy:kg}
        & $16.7\pm1.6\pm1.3$ \cite{Aubert:2001hs}
        & $19.3^{+3.4+1.5}_{-3.2-0.6}$ \cite{Abe:2001nq} \\
    & $\pi^0K^0$ 
        & $14.6^{+5.9+2.4}_{-5.1-3.3}$ \cite{Cronin-Hennessy:kg}
        & $8.2^{+3.1}_{-2.7}\pm1.2$ \cite{Aubert:2001hs}
        & $16.0^{+7.2+2.5}_{-5.9-2.7}$ \cite{Abe:2001nq} \\
    & $\eta K^0$ 
        & $0.0^{+3.2}_{-0.0} \; (<9.3)$ \cite{Richichi:1999kj}
        & - & - \\
    & $\eta' K^0$ 
        & $89^{+18}_{-16}\pm9$ \cite{Richichi:1999kj}
        & $42^{+13}_{-11}\pm4$ \cite{Aubert:2001zf}
        & $55^{+19}_{-16}\pm8$ \cite{Abe:2001pf} \\
    & $\pi^-K^{*+}$ 
        & $22^{+8+4}_{-6-5}$ \cite{Jessop:1999cv} 
        & - 
        & $26.0\pm8.3\pm3.5$ \cite{belle0115} \\
    & $\eta K^{*0}$ 
        & $13.8^{+5.5}_{-4.6}\pm1.6$ \cite{Richichi:1999kj}
        & $19.8^{+6.5}_{-5.6}\pm1.7$ \cite{Aubert:2001ye}
        & $21.2^{+5.4}_{-4.7}\pm2.0$ \cite{Tajima:2001qp,belle0137} \\
    & $\eta' K^{*0}$ 
        & $7.8^{+7.7}_{-5.7} \; (<24)$ \cite{Richichi:1999kj} 
        & - & - \\
    & $K^+ \rho^-$ 
        & $16.0^{+7.6}_{-6.4}\pm2.8 \; (<32)$ \cite{Jessop:2000bv} 
        & - 
        & $15.8^{+5.1+1.7}_{-4.6-3.0}$ \cite{belle0115} \\
    & $K^0 \omega$ 
        & $10.0^{+5.4}_{-4.2}\pm1.4 \; (<21)$ \cite{Jessop:2000bv}
        & $6.4^{+3.6}_{-2.8}\pm0.8 \; (<13)$ \cite{Aubert:2001zf}
        & - \\
    & $K^0 \phi$
        & $5.4^{+3.7}_{-2.7}\pm0.7 \; (<12.3)$ \cite{Briere:2001ue}
        & $8.1^{+3.1}_{-2.5}\pm0.8$ \cite{Aubert:2001zd}
        & $8.9^{+3.4}_{-2.7}\pm1.0$ \cite{Tajima:2001qp} \\
\hline\hline
\end{tabular}
\end{center}
\vspace{0.5cm}
\end{table}

In Table \ref{tab:dS0}, the values of $|t| \simeq |T| = 2.7 \pm 0.6$ and $|p|
\simeq |P| = 0.72 \pm 0.14$ for the $\pi^+\pi^-$ decay mode are based on the
detailed analysis in Ref.\ \cite{Luo:2001ek}.  Here amplitudes are defined such
that their squares give $B^0$ branching ratios in units of $10^{-6}$.  In
estimating $\cB (B^+ \to \pi^+ \pi^0)$ from $T$, we take into account the
lifetime difference between $B^+$ and $B^0$, $\tau_{B^+}/\tau_{B^0} = 1.068 \pm
0.016$ \cite{lifetime} and assume a constructively interfering amplitude $c
\simeq 0.1t$.  The branching ratio thus computed is $\simeq 4.7 \times
10^{-6}$, consistent with the averaged data.  The penguin contribution to $\cB
(B^+ \to K^+ {\ol K}^0)$ is then about $0.55 \times 10^{-6}$.
 
The magnitude of $|p'|^2$ can be directly obtained from the $\pi^+ K^0$ decay
mode to have a central value $\sim 17.2$.  This result is used to compute
$|p|^2$ using the relation $|p/p'|^2 = |\vtd / \vts|^2 \simeq 0.032$, giving
the number quoted above from Ref.\ \cite{Luo:2001ek}.  Here the bounds $0.66
\le |\vtd /\lambda \vts| = |1 - \rho - i \eta| \le 0.96$ on parameters of the
CKM matrix are taken from the analysis of Ref.\ \cite{StA}.

The contributions of $|t'|^2$ are estimated using the relation $|t'/t|^2 =
|\vus / \vud|^2 |f_K/f_\pi|^2 \simeq 0.076$.  We use \cite{PDG} $f_\pi = 130.7$
MeV, $f_K = 159.8$ MeV, $\vus = 0.2205$, and $\vud \simeq 1 - \vus^2/2$.  It
should be noted that the lifetime difference has to be taken into account when
going from $B^0$ to $B^+$ decays.  For $|\Delta S|=1$ decays, the presence of a
substantial electroweak penguin contribution in $c'$ means that one cannot
simply take $c'/t' = 0.1$ as in the $\Delta S = 0$ decays, but must consider
the relative magnitude and weak phase of the electroweak penguin and tree
terms, as in Refs.\ \cite{GR2001,NR}.  Predictions of the branching ratios for
$\pi K$ modes other than $\pi^+ K^0$ depend on both CKM phases and on
final-state phases, which are not yet measured but are likely to be small
\cite{BBNS}.  Extraction of CKM phases from the $\pi K$ modes is a rich area
which we do not address in the present paper.

Two new measurements of the $\pi^+ \rho^0$ and $\pi^{\pm} \rho^{\mp}$ decay
modes are reported in Ref.\ \cite{Gao:2001ce}.  The measurement in the latter
mode does not distinguish between the two final states, while the former
contains a possible penguin contribution.  If we assume $p_V = - p_P$, then
$A(B^+ \to \pi^+ \rho^0) \simeq -\frac{1}{\sqrt{2}}(t_V+c_P-2p_P)$, while
$A(B^+ \to \pi^+ \omega) \simeq \frac{1}{\sqrt{2}}(t_V+c_P+2s_P)$.  Thus,
neglecting the $s_P$ and $c_P$ contributions as in Ref.\ \cite{VPUP}, we may
use $\cB(B^+ \to \pi^+ \omega)$ to estimate the $|t_V|^2$ contribution,
obtaining $(7.9 \pm 1.8) \times 10^{-6}$.  (If we had neglected the penguin
contribution in $B^+ \to \pi^+ \rho^0$ and averaged its branching ratio with
that of $B^+ \to \pi^+ \omega$ we would have obtained instead $(8.8 \pm 1.6)
\times 10^{-6}$, not very different.)  We shall return to the possibility of a
measurable difference between the $\pi^+ \rho^0$ and $\pi^+ \omega$ modes in
Sec.\ \ref{sec:tpi}.

The inferred $|t_V|^2$ contribution to $\cB(B^0 \to \pi^+ \rho^-)$ (neglecting
$c_P$) is $(14.7 \pm 3.3) \times 10^{-6}$, or approximately half of $\cB(B^0
\to \pi^{\pm} \rho^{\mp}) = (25.8 \pm 4.5) \times 10^{-6}$.  This leaves a
contribution of $\cB(B^0 \to \pi^- \rho^+) = (11.1 \pm 5.6) \times 10^{-6}$ to
be supplied by $|t_P|^2$, if we neglect penguin contributions.  A value of
$|t_P|^2$ comparable to $|t_V|^2$, but with large errors, thus is allowed by
present data.  A better measurement of $\cB(B^0 \to \pi^{\pm} \rho^{\mp})$ is
needed to reduce the uncertainty.  The magnitude of $t_P$ is of particular
interest because of the possibility that the smaller $|\Delta S| = 1$ amplitude
$t'_P$, related to $t_P$ by flavor SU(3), could contribute to a rate difference
between $B^+ \to \eta K^{*+}$ and $B^0 \to \eta K^{*0}$ (Sec.\ VI).

We take into account SU(3) breaking in estimating $t'_{V,P}$ by noting the
meson to which the current gives rise: pseudoscalar in $t'_V$ and vector in
$t'_P$.  Thus, we have $|t'_V/t_V|^2 = |\vus/\vud|^2 |f_K/f_\pi|^2$ and
$|t'_P/t_P|^2 = |\vus/\vud|^2 |f_{K^*}/f_\rho|^2$.  We estimate $f_{K^*}/
f_\rho = 1.04 \pm 0.02$ using standard kinematic factors (see Appendix) and
branching ratios for $\tau \to \rho \nu_\tau$ and $\tau \to K^* \nu_\tau$
quoted in Ref.\ \cite{PDG}.

\section{Penguin and electroweak penguin amplitudes \label{sec:pen}}

\subsection{$B \to \eta' K$ decays}

The decays $B^+ \to \eta' K^+$ and $B^0 \to \eta' K^0$ have quite large
branching ratios.  A large fraction of the amplitudes are contributed by
penguin ($p'$) terms, but these are not sufficient.  One must include also
singlet penguin contributions, as introduced in Refs.\ \cite{Dighe:1995gq} and
\cite{DGReta}.

Neglecting $t'$ contributions (to be discussed below), the branching ratios of
$\eta' K^+$ and $\eta' K^0$ modes should have a ratio roughly equal to the
lifetime ratio.  Averaging these two sets of data, we obtain $\cB (B^0 \to
\eta' K^0) \simeq (65.8 \pm 5.2) \times 10^{-6}$, whose central value implies
$(8/3) |s'|^2 \simeq 10.2$ for constructive interference and $41.6$ for no
interference between $p'$ and $s'$.  The corresponding average numbers for $B^+
\to \eta' K^+$ can thus be obtained by the lifetime ratio: \eg, $\cB (B^+ \to
\eta' K^+) \simeq (70.3 \pm 5.5) \times 10^{-6}$.  When $s'$ and $p'$ interfere
constructively, one needs a relatively small value of $s' \simeq 0.49 p'$ to
obtain the observed branching ratios.

\subsection{$B \to K \phi$ decays}

The branching ratios $\cB(B^+ \to K^+ \phi)$ and $\cB(B^0 \to K^0 \phi)$, when
compared with the $p'_P$ contributions, suggest a destructively interfering
$s'_P$.  We associate its contribution with the electroweak penguin component
rather than the $S'_P$ amplitude, which would involve a violation of the
Okubo-Iizuka-Zweig rule unusual for $\omega$ and $\phi$ mesons.

The average of the charged and neutral $B \to K \phi$ modes $\cB (B^+ \to K^+
\phi) = (7.8\pm1.0) \times 10^{-6}$ and $\cB (B^0 \to K^0 \phi) = (7.3 \pm 0.9)
\times 10^{-6}$ are used to extract $s'_P$.  The result is $s'_P/p'_P =
-0.20\pm0.11$, consistent with the result found in Ref.\ \cite{VPUP} (see Table
III there) and with the predictions of Ref.\ \cite{EWPp}.  However, better
measurements of these decay modes and of the mode $B^+ \to \pi^+ K^{*0}$
providing $|p'_P|$ would be worthwhile to confirm the result.

\subsection{$B \to K \omega$ decays}

Electroweak penguin terms arise in $B \to K \omega$ from $c'_P$ and $s'_P$
amplitudes, leading to an overall contribution $+ \frac{1}{3}P'_{EWP}/\sqrt{2}
\simeq -0.20 p'_V/\sqrt{2}$ to each amplitude.  Thus, as in $B \to K \phi$
decays, the electroweak penguin amplitude reduces the contribution of the
dominant penguin amplitude to the rate by about 30\%, and one has the
predictions
\begin{equation}
\cB(B^+ \to K^+ \omega) \simeq \frac{1}{2} \cB(B^+ \to K^+ \phi)
= (3.9 \pm 0.5) \times 10^{-6}~,
\end{equation}
\begin{equation}
\cB(B^0 \to K^0 \omega) = \frac{1}{2} \cB(B^0 \to K^0 \phi) =
(3.7 \pm 0.5) \times 10^{-6}~.
\end{equation}
The former result could be significantly modified by tree-penguin interference,
as noted in Ref.\ \cite{VPUP} and as we shall see in Sec.\ VI.

\section{Rates and $CP$ asymmetries in $B^+ \to \pi^+ (\eta,\eta')$
 \label{sec:pie}}

The decays $B^+ \to \pi^+ \eta$ and $B^+ \to \pi^+ \eta'$ could be detectable
at present levels of sensitivity.  Measurements of the branching ratios and
$CP$ asymmetries of these modes can provide information on strong and weak
phases and on the relative importance of singlet amplitude contributions, which
are estimated using $s'$ in the $\eta' K^+$ mode as discussed above.

We shall give an illustrative example of the possibilities for large rates and
$CP$ asymmetries in $B^+ \to \pi^+ \eta$ and $B^+ \to \pi^+ \eta'$ decays.  We
shall assume that the singlet amplitude $s$ interferes constructively with $p$.
Their electroweak phases are likely to be the same, and a quite modest $s'$
interfering constructively with $p'$ in the decays $B \to \eta' K$ can account
for the observed rate.  We thus take $s/p = s'/p' = 0.49$, leading to the
entries on column (a) of Table II.

Using flavor SU(3) to estimate $p$ from the dominant amplitude $p'$ in $B^+ \to
\pi^+ K^0$ and $t+c$ as mentioned earlier, we then reconstruct the $B^+ \to
\pi^+ (\eta,\eta')$ amplitudes as follows:
\bea
A(B^+ \to \pi^+ \eta) &=&
  -\left( 1.77 e^{i \gamma} + 1.06 e^{-i \beta} e^{i \delta} \right)
  ~, \nonumber \\
A(B^- \to \pi^- \eta) &=&
  -\left( 1.77 e^{-i \gamma} + 1.06 e^{i \beta} e^{i \delta} \right)
  ~, \nonumber \\
A(B^+ \to \pi^+ \eta') &=&
  1.25 e^{i \gamma} + 1.19 e^{-i \beta} e^{i \delta} ~, \nonumber \\
A(B^- \to \pi^- \eta') &=&
  1.25 e^{-i \gamma} + 1.19 e^{i \beta} e^{i \delta} ~,
\eea
where $\beta$ and $\gamma$ are CKM phases, $\delta$ is a relative strong phase
between the penguin and tree amplitudes, and amplitudes are defined such that
their squares give branching ratios in units of $10^{-6}$.

The $CP$ rate asymmetries
\begin{equation}
A(f) \equiv \frac{\cB(B^- \to \bar f) - \cB(B^+ \to f)}
{\cB(B^- \to \bar f) + \cB(B^+ \to f)}
\end{equation}
and the $CP$-averaged branching ratios
\begin{equation}
\overline{\cB}(f) \equiv \frac{\cB(B^- \to \bar f) + \cB(B^+ \to f)}{2}
\end{equation}
then are found to be
\begin{equation}
A(\pi^+ \eta) = \frac{- 0.88 \sin \delta \sin \alpha}{1 - 0.88 \sin \delta \sin
\alpha}~~,~~~
A(\pi^+ \eta') = \frac{- \sin \delta \sin \alpha}{1 - \sin \delta \sin
\alpha},
\end{equation}
\begin{equation}
\overline{\cB}(\pi^+ \eta) = (4.3 \times 10^{-6})(1 - 0.88 \cos \delta
\cos \alpha)~~,~~~
\overline{\cB}(\pi^+ \eta')=(3.0 \times 10^{-6})(1 - \cos \delta \cos \alpha).
\end{equation}
Measurement of both $CP$ asymmetries and branching ratios would allow one to
obtain values of $\delta$ and $\alpha = \pi - \beta - \gamma$, given our
assumption about $s/p$.

\section{Tree-penguin interference \label{sec:tpi}}

\subsection{$B \to \eta' K$ decays}

The central values of the measured rates for $B^+ \to \eta' K^+$ and $B^0 \to
\eta' K^0$ are roughly $1.5 \sigma$ away from each other.  One can attribute
part of this difference to a contribution the tree amplitude in the former
mode, if the tree and penguin amplitudes happen to interfere constructively.
We estimate the $|t'|^2$ term to contribute an amount $0.10 \times 10^{-6}$ to
the branching ratio (see Table III), which by itself would be insignificant.
However, with fully constructive interference with the $p'$ and $s' \simeq 0.49
p'$ terms, we would have
\begin{equation}
\cB(B^+ \to \eta' K^+) = \left( 70.2 + 0.10 + 2 \sqrt{(70.2)(0.10)} \right)
\times 10^{-6} = 75.7 \times 10^{-6}~.
\end{equation}
Thus, in order to demonstrate such interference, one has to conclusively
establish the $B^+ \to \eta' K^+)$ branching ratio with an error of less than a
couple of parts in $10^6$.  At present the errors on the branching ratios are
still too large to give a conclusive answer to whether $t'$ plays an important
role here.

\subsection{$B \to \eta K^*$ decays}

The results for $\cB (B^+ \to \pi^+ K^{*0})$ give $|p'_P|^2 \simeq 12.2 \times
10^{-6}$, implying $\cB (B^+ \to \eta K^{*+}) = 16.2 \times 10^{-6}$ and $\cB(
B^0 \to \eta K^{*0}) = 15.2 \times 10^{-6}$.  Both experimental values are a
bit more than $1 \sigma$ above these predictions.  The question was raised in
Ref.\ \cite{VPUP} whether tree-penguin interference could be responsible for
the slightly higher $\eta K^{*+}$ branching ratio.  The $t'_P$ contribution
here is related to $t_P$ inferred from $B^0 \to \pi^- \rho^+$ by the ratio
$|\vus / \vud|^2 |f_{K^*}/f_{\rho}|^2 \tau_{B^+}/\tau_{B^0} \simeq 0.059$.
With maximal constructive interference we could have a modest enhancement:
\begin{equation}
\cB(B^+ \to \eta K^{*+}) = \left( 16.2 + 0.22 + 2 \sqrt{(16.2)(0.22)}
\right) \times 10^{-6} = 20.2 \times 10^{-6}~,
\end{equation}
To see such an effect, as for $B \to \eta' K$ decays, it would be necessary to
achieve an error on branching ratios of a couple of parts in $10^6$.

Ignoring the contribution from $t'_P$, charged and neutral modes are predicted
to have the same rates.  Taking the average of the current data, we obtain $\cB
(B^+ \to \eta K^{*+}) \simeq (20.3 \pm 3.1) \times 10^{-6}$ and $\cB (B^0 \to
\eta K^{*0}) \simeq (19.0 \pm 2.9) \times 10^{-6}$.  Therefore, at the present
level of sensitivity there is no indication of significant effects due to the
interference of the $t'_P$ amplitude with the dominant penguin contribution.
These data would favor a slightly larger penguin contribution than extracted
from the $\pi^+ K^{*0}$ mode.

\subsection{$B \to \omega K$ decays}

We mentioned above the possibility of tree-penguin interference in $B^+ \to
\omega K^+$.  To give one example of such effects, let us recall the assumption
$p'_V = - p'_P$ but assume the signs of $t'_P$ and $t'_V$ are the same.  Then
if one has constructive interference in $B^+ \to \eta K^{*+}$ as suggested
above, one would have \textit{destructive} interference in $B^+ \to \omega
K^+$.  The $t'_V$ contribution here is related to $t_V$ in $B^+ \to \omega
\pi^+$ by $|t'_V/t_V|^2 = |\vus / \vud|^2 |f_K / f_{\pi}|^2 \simeq 0.076$.  In
the case of maximal destructive interference one would have
\begin{equation}
\cB(B^+ \to \omega K^+) = \left( 3.9 + 0.6 - 2 \sqrt{(3.9)(0.6)} \right)
\times 10^{-6} = 1.4 \times 10^{-6},
\end{equation}
a significant effect.

\subsection{$B^0 \to \pi^- K^{*+}$ and $B^0 \to K^+ \rho^-$ decays}

The signs of tree-penguin interference terms in the decays $B^0 \to \pi^-
K^{*+}$ and $B^0 \to K^+ \rho^-$ are correlated with those in $B^+ \to K^+
\omega$.  If the interference is destructive in $B^+ \to K^+ \omega$, it will
also be destructive in $B^0 \to K^+ \rho^-$, since both processes involve the
combination $p'_V + t'_V$.  If $t'_P$ and $t'_V$ have the same sign (as is
likely), but if $p'_P$ and $p'_V$ are equal and opposite (as has been
proposed), one then expects constructive tree-penguin interference in $B^0 \to
\pi^- K^{*+}$.  This pattern was noted in Refs.\ \cite{VPUP} and \cite{B2K}.

In the cases of maximal interference in the directions suggested, one would
then have
\begin{equation}
\cB(B^0 \to \pi^- K^{*+}) = \left( 11.4 + 0.6 + 2 \sqrt{(11.4)(0.6)} \right)
\times 10^{-6} = 17.3 \times 10^{-6},
\end{equation}
consistent with the experimental branching ratio of $(23.8 \pm 6.1)
\times 10^{-6}$, but also
\begin{equation}
\cB(B^0 \to K^+ \rho^-) = \left( 11.4 + 1.1 - 2 \sqrt{(11.4)(1.1)} \right)
\times 10^{-6} = 5.4 \times 10^{-6},
\end{equation}
which is well below the experimental branching ratio of $(15.9 \pm 4.4)
\times 10^{-6}$.  In each case the deviation from pure penguin dominance
amounts to $6 \times 10^{-6}$, so measurement of each of these branching
ratios with an error of no more than $2 \times 10^{-6}$ should be enough
to see whether the interference terms form a consistent pattern, or indeed
are present at all.

\subsection{$B^+ \to \pi^+ \rho^0$ and $B^+ \to \pi^+ \omega$ decays}

More precise measurements for the $B^+ \to \pi^+ \rho^0$ and $B^+ \to \pi^+
\omega$ modes could help to determine whether there is a difference between
their branching ratios, which would be ascribed to contributions of the $p_P$
and/or $s_P$ amplitudes.  The chance of a detectable $s_P$ contribution to $B^+
\to \pi^+ \phi$, for which BaBar has presented an upper bound
\cite{Aubert:2001zd}, is remote, as one sees from the predicted branching ratio
of about $2 \times 10^{-8}$ in Table II.  Consequently, one would most likely
ascribe a difference to constructive tree-penguin interference, which would be
consistent with the pattern mentioned earlier \cite{VPUP,B2K}, leading to a
prediction
\begin{equation}
\cB(B^+ \to \pi^+ \rho^0) = \left( 7.9 + 0.8 + 2 \sqrt{(7.9)(0.8)} \right)
\times 10^{-6} = 13.6 \times 10^{-6}
\end{equation}
As in previous cases, the effects of maximal interference amount to a change
in the predicted branching ratio of a few parts in $10^6$.

\section{Further singlet amplitude contributions \label{sec:sin}}

We have already noted in Sec.\ IV the importance of the singlet contribution
$s'$ in the decays $B \to \eta' K$.  However, no such contribution is yet
called for in $B \to PV$ decays.  Here we show how to demonstrate its presence.

% This is Figure 1
\begin{figure}[ht]
\vspace{0.5cm}
\centerline{\includegraphics[height=4in]{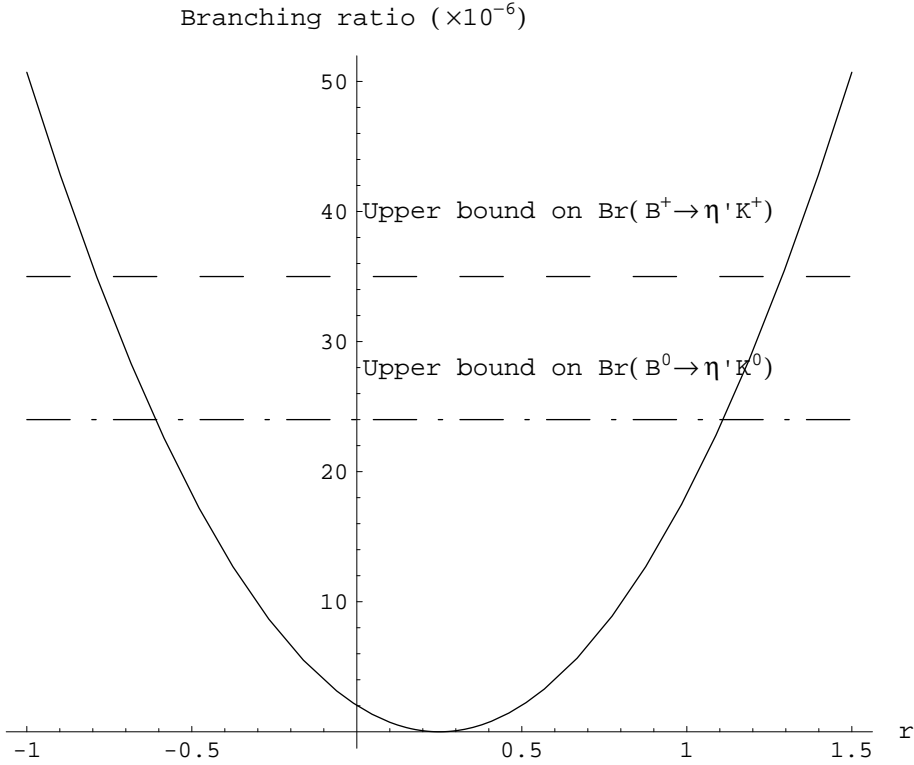}}
\vspace{0.5cm}
\tighten{
\caption[]{\it The branching ratios of $B \to \eta' K^*$ for varying
  $s'_V$ related to $p'_P$ by the parameter $-1 \leq r \leq 1.5$.}
\label{fig:plot1}}
\end{figure}

A contribution from the singlet amplitude $s'_V$ has to come from the
comparison between the $\eta K^*$ and $\eta' K^*$ modes.  If we neglect $t'_P$,
as suggested from the above analysis, and $c'_V$, as suggested by the hierarchy
in the amplitudes, we can assume $s'_V = r p'_P$ and get
\be
\cB (B^+ \to \eta' K^{*+}) 
     \simeq \cB (B^0 \to \eta' K^{*0}) \frac{\tau_{B^0}}{\tau_{B^+}}
     = \frac16 (1 - 4r)^2 p'^2_P ~,
\ee
where $p'_P$ is the penguin amplitude for the charged modes.  Fig.\ 
\ref{fig:plot1} shows the branching ratio of $B \to \eta' K^{*}$ as a parabolic
function of $r$ with a minimum at $r=1/4$.  To avoid confusion, we only plot
the one for $B \to \eta' K^{*+}$ as the difference is tiny in the range of the
plot.  The dashed and dash-dotted lines give the current upper bounds on the
branching ratios of the $\eta' K^{*+}$ and $\eta' K^{*0}$ modes, respectively.
Observation of these modes with branching ratios significantly different from
$\sim 2 \times 10^{-6}$ would provide conclusive evidence for the singlet
contribution $s'_V$.  We note that $\cB (B^+ \to \eta' K^{*+})$ by itself is
unable to distinguish between $r$ and $r' \equiv \frac{1}{2} - r$, so if this
branching ratio is consistent with $\sim 2 \times 10^{-6}$, that does not yet
rule out the possibility of a singlet term with $s'_V/p'_P \simeq 1/2$.  This
is just the value of $s'/p'$ which would accommodate the decays $B \to \eta'
K$.

\section{Summary \label{sec:sum}}

New data on $B$ decays to pairs of light mesons are shedding light on a number
of interesting questions.  We have shown that the penguin contribution in the
decay $B^+ \to \pi^+ K^{*0}$ is only a bit smaller than that contributing to $B
\to \pi K$ decays.  Although a similar penguin contribution occurs in $B \to K
\phi$ decays, it is partially cancelled by an electroweak penguin contribution,
leading to a 30\% reduction in rate in accord with predictions \cite{EWPp}.  A
similar cancellation is expected in the decays $B \to K \omega$.

The prospects for observing $B^+ \to \pi^+ \eta$ and $B^+ \to \pi^+ \eta'$,
suggested as promising modes for direct $CP$ rate asymmetries
\cite{BSR,Dighe:1995gq}, are excellent.  Branching ratios of a few parts in
$10^6$ are expected.  By studying both rates and $CP$ asymmetries, one can
determine both the relative strong phases of penguin and tree amplitudes and
the weak phase $\alpha$.

Tree-penguin interference can be studied by comparing $B^+$ and $B^0$ branching
ratios for processes such as $B \to \eta' K$, $B \to \eta K^*$, and $B \to K
\omega$.  Anticipated differences in branching ratios in these three cases
could be as large as several parts in $10^6$, but are unlikely to be more.
Other processes which can be examined for this interference include the decays
$B^0 \to \pi^- K^{*+}$, $B^0 \to K^+ \rho^-$, and a comparison of $B^+ \to
\pi^+ \omega$ and $B^+ \to \pi^+ \rho^0$.  Present data are not yet at the
required level of accuracy, but will be so soon, providing valuable information
on the products $\cos \gamma \cos \delta$ ($|\Delta S| = 1$ decays) and
$\cos \alpha \cos \delta$ ($\Delta S = 0$ decays).

Although a flavor-singlet penguin contribution is needed in describing $B \to
\eta' K$, no such amplitude is called for yet in $B \to \eta' K^*$.  We have
shown that significant deviations of the branching ratio for this process (for
both charged and neutral $B$'s) from $2 \times 10^{-6}$ would provide evidence
for such a term.  However, a branching ratio equal to this value does not yet
rule out a singlet term.

\begin{acknowledgments}
We thank H. J. Lipkin for helpful discussions, and D. Hitlin and M. Nakao for
guidance with respect to data.  This work was supported in part by the
U. S. Department of Energy through Grant Nos.\ DE-FG02-90ER-40560
and W-31109-ENG-38.
\end{acknowledgments}

\appendix*
\section{Decay constant calculations}

We define the decay constant of a vector meson $V (=u {\ol q})$ through the
matrix element between one particle and vacuum of the vector current $V_\mu$:
$\langle 0 |V_\mu| V(p) \rangle = m_V f_V \epsilon_\mu(p)$.  The partial width
of the $\tau$ lepton into $V \nu_\tau$ is then
\begin{equation}
\Gamma(\tau \to V \nu_\tau) = \frac{(G_F f_V p^* |V_{uq}|)^2}{4 \pi}
m_\tau \left( 1 + \frac{2 m_V^2}{m_\tau^2} \right) ~,
\end{equation}
where $p^* = (m_{\tau}^2 - m_V^2)/(2 m_{\tau})$ is the magnitude of the c.\ m.\ 
three-momentum of either final particle, and $|V_{uq}| = |\vud|$ for $\rho
\nu_{\tau}$ or $|\vus|$ for $K^* \nu_{\tau}$.  Using \cite{PDG} $\tau_\tau =
(290.6 \pm 1.1)$ fs, $\cB (\tau \to \rho \nu_\tau) = (25.1 \pm 0.3)\%$, and
$\cB (\tau \to K^* \nu_\tau) = (1.29 \pm 0.05)\%$,
we find $f_\rho = 208$ MeV, $f_{K^*} = 217$ MeV, and $f_{K^*}/ f_\rho = 1.04
\pm 0.02$.


\begin{thebibliography}{99}

\bibitem{VPUP}
M.~Gronau and J.~L.~Rosner,
%``New information on B decays to charmless V P final states,''
Phys.\ Rev.\ D {\bf 61}, 073008 (2000) [hep-ph/9909478].
%%CITATION = HEP-PH 9909478;%%

\bibitem{HJLP}
H.~J.~Lipkin,
%``The Importance Of The K Eta And K Eta-Prime Decay Modes In Understanding
%Charmed And Other Meson Decays,''
Phys.\ Rev.\ Lett.\ {\bf 46}, 1307 (1981);
%%CITATION = PRLTA,46,1307;%%
%``Interference Effects In K Eta And K Eta-Prime Decay Modes Of Heavy Mesons:
%Clues To Understanding Weak Transitions And $CP$ Violation,''
Phys.\ Lett.\ B {\bf 254}, 247 (1991);
%%CITATION = PHLTA,B254,247;%%
%``Penguins, trees and final state interactions in B decays in broken SU(3),''
\textit{ibid.} B {\bf 415}, 186 (1997) [hep-ph/9710342];
%%CITATION = HEP-PH 9710342;%%
%H.~J.~Lipkin,
%``Fsi Rescattering In B+- Decays Via States With Eta, Eta' Omega And Phi,''
\textit{ibid.} B {\bf 433}, 117 (1998).
%%CITATION = PHLTA,B433,117;%%

\bibitem{BSR}
S.~Barshay, D.~Rein and L.~M.~Sehgal,
%``$CP$ Violating Partial Rate Asymmetries In The Decays B+- $\to$ Eta Pi+-,
%Eta-Prime Pi+-, Eta(C) Pi+-: A K Matrix Analysis,''
Phys.\ Lett.\ B {\bf 259}, 475 (1991).
%%CITATION = PHLTA,B259,475;%%

\bibitem{Dighe:1995gq}
A.~S.~Dighe, M.~Gronau and J.~L.~Rosner,
%``Amplitude relations for $B$ decays involving $\eta$ and $\eta'$,''
Phys.\ Lett.\ B {\bf 367}, 357 (1996) [hep-ph/9509428]; {\it Erratum-ibid.} B
{\bf 377}, 325 (1996).
%%CITATION = HEP-PH 9509428;%%

\bibitem{HHY}
X.-G. He, W.-S. Hou, and K.-C. Yang,
Phys.\ Rev.\ Lett.\ {\bf 83}, 1100 (1999).

\bibitem{HSW} W.-S. Hou, J. G. Smith, and F. W\"urthwein, National Taiwan
University report NTU-HEP-99-25, hep-ex/9910014, submitted to Phys.~Rev.~Lett.

\bibitem{HY}
W.-S. Hou and K.-C. Yang,
Phys.\ Rev.\ D {\bf 61}, 073014 (20000 [hep-ph/9908202].

\bibitem{DGReta}
A.~S.~Dighe, M.~Gronau and J.~L.~Rosner,
%``B decays involving eta and eta' in light of the B $\to$ K eta' process,''
Phys.\ Rev.\ Lett.\ {\bf 79}, 4333 (1997) [hep-ph/9707521].
%%CITATION = HEP-PH 9707521;%%
 
\bibitem{Gronau:1994rj}
M.~Gronau, O.~F.~Hernandez, D.~London and J.~L.~Rosner,
%``Decays of B mesons to two light pseudoscalars,''
Phys.\ Rev.\ D {\bf 50}, 4529 (1994) [hep-ph/9404283].
%%CITATION = HEP-PH 9404283;%%

\bibitem{Gronau:1995hn}
M.~Gronau, O.~F.~Hernandez, D.~London and J.~L.~Rosner,
%``Electroweak penguins and two body B decays,''
Phys.\ Rev.\ D {\bf 52}, 6374 (1995) [hep-ph/9504327].
%%CITATION = HEP-PH 9504327;%%

\bibitem{EWVP}
M.~Gronau,
%``Electroweak Penguin Amplitudes And Constraints On Gamma In Charmless B $\to$
%Vp Decays,''
Phys.\ Rev.\ D {\bf 62}, 014031 (2000).
%%CITATION = PHRVA,D62,014031;%%

\bibitem{GR2001}
M.~Gronau and J.~L.~Rosner,
%``Implications of $CP$ asymmetry limits for B $\to$ K pi and B $\to$ pi pi,''
Phys.\ Rev. D {\bf 65}, 013004 (2002), [hep-ph/0109238].
%%CITATION = HEP-PH 0109238;%%

\bibitem{Jessop:1999cv}
C.~P.~Jessop {\it et al.}  [CLEO Collaboration],
%``Two-Body B Meson Decays to $\eta$ and $\eta^{'}$-Observation of $B\to \eta
%K^{*}$,'' 
CLEO Report No.\ CLEO-CONF 99-13, hep-ex/9908018.
%%CITATION = HEP-EX 9908018;%%

\bibitem{Cronin-Hennessy:kg}
D.~Cronin-Hennessy {\it et al.}  [CLEO Collaboration],
%``Observation Of B $\to$ K+- Pi0 And B $\to$ K0 Pi0, And Evidence For
%B $\to$ Pi+ Pi-,''
Phys.\ Rev.\ Lett.\ {\bf 85}, 515 (2000).
%%CITATION = PRLTA,85,515;%%

\bibitem{Richichi:1999kj}
S.~J.~Richichi {\it et al.}  [CLEO Collaboration],
%``Two-body B meson decays to eta and eta': Observation of B $\to$ eta K*,''
Phys.\ Rev.\ Lett.\ {\bf 85}, 520 (2000) [hep-ex/9912059].
%%CITATION = HEP-EX 9912059;%%

\bibitem{Jessop:2000bv}
C.~P.~Jessop {\it et al.}  [CLEO Collaboration],
%``Study of charmless hadronic B meson decays to pseudoscalar vector
%final states,''
Phys.\ Rev.\ Lett.\  {\bf 85}, 2881 (2000) [hep-ex/0006008].
%%CITATION = HEP-EX 0006008;%%

\bibitem{Briere:2001ue}
R.~A.~Briere {\it et al.}  [CLEO Collaboration],
%``Observation of B $\to$ Phi K and B $\to$ Phi K*,''
Phys.\ Rev.\ Lett.\  {\bf 86}, 3718 (2001) [hep-ex/0101032].
%%CITATION = HEP-EX 0101032;%%

\bibitem{Asner:2001eh}
D.~M.~Asner {\it et al.}  [CLEO Collaboration],
%``Search for B0 $\to$ pi0 pi0 decay,''
Cornell University Report No.\ CLNS-01/1718, hep-ex/0103040.
%%CITATION = HEP-EX 0103040;%%

\bibitem{Gao:2001ce}
Y.~s.~Gao [CLEO Collaboration],
%``Recent results from CLEO collaboration,''
Southern Methodist University Report No.\ SMU-HEP-01-09, hep-ex/0108005, to
appear in proceeding of the International Conference on Flavor Physics
(ICFP2001).
%%CITATION = HEP-EX 0108005;%%

\bibitem{Aubert:2000vr}
B.~Aubert {\it et al.}  [BaBar Collaboration],
%``Measurements of charmless three-body and quasi-two-body B decays,''
SLAC Report No.\ SLAC-PUB-8537, hep-ex/0008058, submitted to the XXX
International Conference on High Energy Physics, Osaka, Japan, July 27 --
August 2, 2000.
%%CITATION = HEP-EX 0008058;%%

\bibitem{Aubert:2001zd}
B.~Aubert {\it et al.}  [BaBar Collaboration],
%``Measurement of the decays B $\to$ Phi K and B $\to$ Phi K*,''
Phys.\ Rev.\ Lett.\  {\bf 87}, 151801 (2001) [hep-ex/0105001].
%%CITATION = HEP-EX 0105001;%%

\bibitem{Aubert:2001hs}
B.~Aubert {\it et al.}  [BaBar Collaboration],
%``Measurement of branching fractions and search for CP-violating charge
%asymmetries in charmless two-body B decays into pions and kaons,''
Phys.\ Rev.\ Lett.\  {\bf 87}, 151802 (2001) [hep-ex/0105061].
%%CITATION = HEP-EX 0105061;%%

\bibitem{Aubert:2001ye}
B.~Aubert {\it et al.}  [BaBar Collaboration],
%``Measurement of the exclusive branching fractions $B^0 \to \eta
%K^{*0}$ and $B^+ \to \eta K^{*+}$,''
SLAC Report No.\ SLAC-PUB-8914, hep-ex/0107037.  Submitted to the International
Europhysics Conference on High Energy Physics, 12--18 July 2001, Budapest,
Hungary.
%%CITATION = HEP-EX 0107037;%%

\bibitem{Aubert:2001zf}
B.~Aubert {\it et al.}  [BaBar Collaboration],
%``Measurements of the branching fractions of exclusive charmless $B$
%meson decays with $\eta^\prime$ or $\omega$ mesons,''
Phys.\ Rev.\ Lett.\ {\bf 87}, 221802 (2001) [hep-ex/0108017].
%%CITATION = HEP-EX 0108017;%%

\bibitem{Aubert:2001ap}
B.~Aubert {\it et al.}  [BaBar Collaboration],
%``Measurement of the branching fraction for B+ $\to$ K*0 pi+,''
SLAC Report No.\ SLAC-PUB-8981, hep-ex/0109007, submitted to the 9th
International Symposium on Heavy Flavor Physics, Pasadena, California, 10--13
September 2001.
%%CITATION = HEP-EX 0109007;%%

\bibitem{Dallapiccola}
C.~Dallapiccola [BaBar Collaboration], talk presented at the 9th International
Symposium on Heavy Flavor Physics, Pasadena, California, 10--13 September 2001.

\bibitem{Abe:2001nq}
K.~Abe {\it et al.}  [Belle Collaboration],
%``Measurement of branching fractions for B $\to$ pi pi, K pi and K K
%decays,''
Phys.\ Rev.\ Lett.\ {\bf 87}, 101801 (2001) [hep-ex/0104030].
%%CITATION = HEP-EX 0104030;%%

\bibitem{Bozek:2001xd}
A.~Bo\.{z}ek  [Belle Collaboration],
%``Charmless B decays involving vector mesons in Belle,''
hep-ex/0104041; in \textit{BCP4:  International Workshop on B Physics and
CP Violation}, Ise-Shima, Japan, 19--23 February 2001, edited by T. Ohshima
and A. I. Sanda (World Scientific, River Edge, NJ, 2001), p.\ 81.
%%CITATION = HEP-EX 0104041;%%

\bibitem{Abe:2001pf}
K.~Abe {\it et al.}  [Belle Collaboration],
%``Measurement of the branching fraction for B $\to$ eta' K and search
%for B $\to$ eta' pi+,''
Phys.\ Lett.\ B {\bf 517}, 309 (2001) [hep-ex/0108010].
%%CITATION = HEP-EX 0108010;%%

\bibitem{Tajima:2001qp}
H.~Tajima  [Belle Collaboration],
%``Belle B physics results,''
KEK Preprint 2001-136, hep-ex/0111037,
contributed to the Proceedings of the XX International Symposium on Lepton and
Photon Interactions at High Energies, July 23--28, 2001, Rome, Italy.
%%CITATION = HEP-EX 0111037;%%

\bibitem{belle0115}
K.~Abe {\it et al.}  [Belle Collaboration],
BELLE-CONF-0115 (2001), contributed to the Proceedings of the XX International
Symposium on Lepton and Photon Interactions at High Energies, July 23--28,
2001, Rome, Italy.

\bibitem{belle0137} 
K.~Abe {\it et al.} [Belle Collaboration], 
BELLE-CONF-0137 (2001), submitted to the 9th International Symposium on Heavy
Flavor Physics, Pasadena, California, 10--13 September 2001.

\bibitem{Luo:2001ek}
Z.~Luo and J.~L.~Rosner,
%``Information on B $\to$ pi pi provided by the semileptonic process B
%$\to$ p l nu,''
Enrico Fermi Institute Report No.\ EFI 01-28,hep-ph/0108024, to appear in
Phys.\ Rev.\ D.
%%CITATION = HEP-PH 0108024;%%

\bibitem{lifetime}
K.~Osterberg, talk presented at the International European Conference on
High-Energy Physics, Budapest, Hungary, 12--18 July 2001, to appear in the
Proceedings.

\bibitem{StA} 
J.~L.~Rosner,
%``The standard model in 2001,''
Enrico Fermi Institute Report No.\ EFI 01-34, hep-ph/0108195; based on five
lectures at the 55th Scottish Universities' Summer School in Particle Physics,
St.\ Andrews, Scotland, August 7--23, 2001, to be published in the Proceedings
by the Institute of Physics (U.K.).
%%CITATION = HEP-PH 0108195;%%

\bibitem{PDG}
D.~E.~Groom \textit{et al.} [Particle Data Group], Eur.\ Phys.\ J. C {\bf 15},
1 (2000).
%%CITATION = EPHJA,C15,1;%%

\bibitem{NR}
M.~Neubert and J.~L.~Rosner,
%``New bound on gamma from B+- $\to$ pi K decays,''
Phys.\ Lett.\ B {\bf 441}, 403 (1998) [hep-ph/9808493];
%%CITATION = HEP-PH 9808493;%%
%``Determination of the weak phase gamma from rate measurements in  B+- $\to$
%pi K, pi pi decays,''
Phys.\ Rev.\ Lett.\  {\bf 81}, 5076 (1998) [hep-ph/9809311];
%%CITATION = HEP-PH 9809311;%%
M.~Neubert,
%``Model-independent analysis of B $\to$ pi K decays and bounds on the weak
%phase gamma,''
JHEP {\bf 9902}, 014 (1999) [hep-ph/9812396];
%%CITATION = HEP-PH 9812396;%%
M.~Gronau, D.~Pirjol and T.~M.~Yan,
%``Model-independent electroweak penguins in B decays to two  pseudoscalars,''
Phys.\ Rev.\ D {\bf 60}, 034021 (1999) [hep-ph/9810482].
%%CITATION = HEP-PH 9810482;%%

\bibitem{BBNS}
M.~Beneke, G.~Buchalla, M.~Neubert and C.~T.~Sachrajda,
%``QCD factorization in B $\to$ pi K, pi pi decays and extraction of
%Wolfenstein parameters,''
Nucl.\ Phys.\ B {\bf 606}, 245 (2001) [hep-ph/0104110].
%%CITATION = HEP-PH 0104110;%%

\bibitem{EWPp}
R.~Fleischer,
%``Electroweak Penguin effects beyond leading logarithms in the B
%meson decays B- $\to$ K- Phi and B- $\to$ pi- anti-K0,''
Z.\ Phys.\ C {\bf 62}, 81 (1994);
%%CITATION = ZEPYA,C62,81;%%
N.~G.~Deshpande and X.~G.~He,
%``Gluonic penguin B decays in Standard and two Higgs doublet Models,''
Phys.\ Lett.\ B {\bf 336}, 471 (1994) [hep-ph/9403266].
%%CITATION = HEP-PH 9403266;%%

\bibitem{B2K}
J.~L.~Rosner, in Proceedings of Beauty 2000, Kibbutz Maagan,
Israel, September 13--18, 2000, edited by S.~Erhan, Y.~Rozen, and P.~E.~
Schlein, Nucl.\ Inst.\ Meth.\ A {\bf 462}, 44--51 (2001).
%CITATION = HEP-PH 0011184;%%

\end{thebibliography}
\end{document}